\documentclass[showpacs,prb,preprint,tightenlines,groupedaddress,amssymb]{revtex4}

\usepackage{graphicx}

\begin{document}


\title{Magnetic Transitions and Ferromagnetic Clusters in 
RuSr$_{2}$(Eu,Ce)$_{2}$Cu$_{2}$O$_{10+\delta }$}
\author{Y. Y. Xue}
\author{B. Lorenz}
\author{D. H. Cao}
\author{C. W. Chu}
\altaffiliation[Also at ]{Lawrence Berkeley National Laboratory, 1 Cyclotron 
Road, Berkeley, CA 94720;
and Hong Kong University of Science and Technology, Hong Kong}
\affiliation{Department of Physics and Texas Center for Superconductivity, 
University of Houston, Houston TX 77204-5002}
\date{\today}

\begin{abstract}
The macroscopic magnetizations of a 
RuSr$_{2}$(Eu$_{0.7}$Ce$_{0.3}$)$_{2}$Cu$_{2}$O$_{10+\delta}$ 
sample were investigated. A ferromagnet-like transition
occurs around $T_{M}$ in the low-field magnetization. Highly nonlinear $M(H)$,
non-Curie-Weiss susceptibility, and slow spin-dynamics, however, were
observed up to $T_{1}$ $\approx$ 2--3 $T_{M}$. In addition, an
antiferromagnet-like differential-susceptibility maximum of Ru appears
around a separate temperature $T_{AM}$ between $T_{1}$ and $T_{M}$. The data are
therefore consistent with a phase-separation model: superparamagnetic
clusters (or short-range spin-orders) are first precipitated from the
paramagnetic matrix below $T_{1}$, followed by an antiferromagnetic
transition of the matrix at $T_{AM}$ and an apparent ferromagnetic (FM) 
transition around $T_{M}$, where the long-range spin-order is established 
in the FM species
imbedded in the matrix.
\end{abstract}

\maketitle

\section{Introduction}

The magnetic structure of the rutheno-cuprates RuSr$_{2}$RCu$_{2}$O$_{8+\delta}$ 
(Ru1212R) and RuSr$_{2}$(R,Ce)$_{2}$Cu$_{2}$O$_{10+\delta}$
(Ru1222R), where R = Gd, Eu, and Y, remains a topic of debate even after 
extensive investigations of
their macroscopic magnetization (M), neutron powder diffraction (NPD), and
nuclear magnetic resonance (NMR) in the last few 
years.\cite{ber,fel,lyn,but,xue,tok} 
Even in the relatively simple case of
Ru1212, a large discrepancy exists. A NPD investigation reported that the
Ru-spins ordered antiferromagnetically (AFM) \textit{along} the $c$-axis with 
a rather stringent upper limit of 0.1 and 0.2~$\mu _{B}$/Ru at H
= 0 and 7 T, respectively, for the ferromagnetic (FM) components of 
Ru.\cite{lyn} The spontaneous magnetization of $\approx 800$~emu/mole at 5~K,
however, leads to a significantly higher value of 0.28~$\mu _{B}$/Ru after a
geometric correction-factor of 2 for the random grain-orientation.\cite{ber}
The difference between the extrapolated zero-field magnetization of 
0.7~$\mu_{B}$/Ru and the NPD estimation of $\approx$~0.2~$\mu _{B}$ at 
7~T is even
larger.\cite{but,lyn} The discrepancy is similarly striking when the NMR
data are included. The zero-field NMR data suggest that the aligned moment, 
$\approx$~1.6~$\mu _{B}$/Ru, should be \textit{perpendicular} to the 
$c$-axis instead, and that the FM component should be a large part of the
moment.\cite{tok} While at least two other NPD studies claim similarly
negligible FM components,\cite{cam,tak} both the large FM moment and the
perpendicular spin-direction were confirmed by independent NMR 
measurements.\cite{kum} These conflicts will be difficult to reconcile, 
in our opinion,
if these rutheno-cuprates are microscopically homogeneous.

The magnetic behavior of Ru1222 appears to be even more 
complicated.\cite{fel,xue} On one hand, the two compounds share the same basic 
features:
well-defined FM-like features in low-field magnetizations around a
transition temperature of $T_{M}$; spontaneous moments far larger than 
0.1 $\mu _{B}$/Ru;\cite{ber,wil,fel} similar Curie constants and FM-like
Curie-Weiss (C-W) temperatures;\cite{but,wil} and significant
nonlinear isothermal $M(H)$ with large extrapolated zero-field moments up to
2--3 $T_{M}$.\cite{but,fel,xue} In particular, when the zero-field NMR
suggests similar large FM components,\cite{tok,fela} a preliminary NPD
investigation observed a negligible ferromagnetic moment in Ru1222Gd as 
well.\cite{kne} The two compounds seem to share an unusual magnetic ground state,
which appears AFM-like in the NPD data but FM-like in the $M(H)$ and NMR
signals. On the other hand, quantitative differences exist between them:
both the nonlinearity in $M(H)$ and the spontaneous magnetization are
significantly larger in Ru1222.\cite{fel,wil}

It is interesting to note that some of the features seem to be
characteristic of magnetic nanoparticles (superparamagnets). As will be
discussed below, the nonlinear $M(H)$ far above $T_{M}$ and the discrepancy
between the NPD and NMR data are very reminiscent of some manganites,
where phase-separations and the formation of ferromagnetic nanoclusters are
well documented. A close re-examination of the magnetic properties of
rutheno-cuprates, therefore, may help to resolve the magnetic structure of
Ru1222. Indeed, our static and dynamic magnetic data of a Ru1222Eu sample
with $T_{M} \approx 65$~K are consistent with a phase-separation model:
ferromagnetic clusters (or short-range spin-orders) are first precipitated
from the paramagnetic matrix below $T_{1} \approx$ 160--180~K, followed by
an AFM transition of the matrix around $T_{AM} \approx 104$~K and an
apparent FM transition around T$_{M}$, which we attribute to the
establishment of long-range FM spin-orders. The model offers a possible
explanation for the conflicting NPD/NMR data as well as the
Josephson-junction-like superconductivity at lower 
temperatures,\cite{lyn,tok,xuea} although detailed structure investigations 
are needed.

\section{Experiment}

Ceramic RuSr$_{2}$(Eu$_{1-x}$Ce$_{x}$)$_{2}$Cu$_{2}$O$_{10+\delta}$ samples
with $0.3 \leq x \leq 0.5$ were synthesized following the standard
solid-state-reaction procedure. Precursors were first prepared by calcining
commercial oxides at 400--900~$^{\circ}$C under flowing O$_{2}$. Mixed
powder with a proper cation-ratio was then pressed into pellets and sintered
at 900~$^{\circ}$C in air for 24~h. The final heat treatment of the
ceramics was done at 1090~$^{\circ}$C for 60~h after repeatedly
sintering at 1000~$^{\circ}$C and regrinding at room temperature.\cite{xue} 
The structure of the samples was determined by powder X-ray
diffraction (XRD) using a Rigaku DMAX-IIIB diffractometer. The XRD pattern
of the samples is similar to that previously reported.\cite{xuea} There are
no noticeable impurity lines in the X-ray diffraction pattern within our
experimental resolution of a few percent. The grain sizes ($\approx$ 
2--20~$\mu$m) of the ceramic samples were measured using a JEOL JSM 6400 scanning
electron microscope (SEM). The magnetizations were measured using a Quantum
Design SQUID magnetometer with an $ac$ attachment. Only the data of a
RuSr$_{2}$(Eu$_{0.7}$Ce$_{0.3}$)$_{2}$Cu$_{2}$O$_{10+\delta }$ sample will be
presented here since the features concerned are rather similar in more than
five samples with various Ce-doping of $0.3 \leq x \leq 0.5$.

\section{Data and Discussion}

\subsection{$dc$ Magnetization and its $H$-Dependence}

The zero-field-cooled magnetization, $M_{ZFC}$, and the field-cooled 
magnetization, 
$M_{FC}$, of the Ru1222Eu sample at 10~Oe are shown in Fig.~\ref{fig:fig1}. Both 
$M_{ZFC}$ and $M_{FC}$ rise FM-like with cooling and 
the separation between $M_{FC}$ and
$M_{ZFC}$ develops below 60 K, phenomena similar to those reported
previously.\cite{fel,wil} Such behavior is typical of magnets with
domain-wall pinning. The inflection point of the $M_{FC}$(10~Oe) is taken as
the magnetic transition temperature $T_{M} \approx 65$~K (shown by the
arrow in Fig.~\ref{fig:fig1}). To consider the effects of possible spin fluctuations, the
observed $M_{FC}/H$ at various $H$ is plotted against $t = T/T_{M}-1$ since the 
spin fluctuations in a second-order transition are expected to
be universal functions of $t$ after a proper $H$ scaling 
(Fig.~\ref{fig:fig2}).\cite{kaul} Two features are readily noticed. First, two additional
transitions, which are too small to be noticed in the linear $M$-scale of Fig.~\ref{fig:fig1}, 
appear far above $T_{M}$ when $H << 0.1$~T.
Similar features have been reported before, and attributed to either
chemical inhomogeneity or a possible phase separation.\cite{fel,xue} 
Their
direct effect on the high-field magnetization, however, seems to be small, 
\textit{i.e.} both $M_{FC}$ and $dM_{FC}/dT$ appear to be smooth functions
of $T$ with no transition-like features at the corresponding temperatures. 
We will therefore not discuss them further. Second, strong nonlinearity, 
\textit{i.e.} an $H$-dependence of $M_{FC}/H$, appears up to $T_{1} \approx$
160--180~K with $H$ between 0.1 and 5~T, where the contributions of the two
additional transitions should be negligible. Similar nonlinearity has been
noticed previously in both Ru1212 and Ru1222 up to $t =$ 1--2 and
has been attributed to spin-fluctuation,\cite{wil} magnetic 
anisotropy,\cite{but} and
rotation of the canting angle.\cite{fel} It should be noted
that the magnetic hysteresis is rather small above $T_{M}$. The separation
between the $H$-increase and the $H$-decrease branches in a $\pm 5$~T 
$M$-$H$ loop is
less than 1\% with $T > 70$~K and $H \geq 500$~Oe. The $M_{FC}$
presented here, therefore, can be treated as the equilibrium magnetization
and discussed in terms of the free energy involved.



Although $H$-dependent $M(H)$'s are a common phenomena in magnets near their
transition temperatures due to fluctuation, its appearance up to $t =$ 
1--2 is highly unusual. In principle, the spin fluctuation near $T_{M}$ is
determined by the competition between the magnetic interaction-energy 
$M_{0}V_{c}H$ and the thermal energy $k_{B}T$, where $M_{0}$ and $V_{c}$
are the aligned moment and the coherent volume, respectively. The
$H$-dependence will be significant only if the magnetic energy is comparable
to or larger than $k_{B}T$. The value of $M_{0}V_{c}$ above $T_{M}$,
however, should decrease rapidly with the increase of $t$ and
should approach the moment $\mu_{0}$ of individual spins outside a narrow
critical region based on the scaling theory.\cite{kaul} Far above $T_{M}$,
the fluctuation-caused nonlinearity should only exist at 
$H \geq k_{B}T/\mu _{0}$. The layered structure of Ru1222 should not change the
conclusion.\cite{note} Although nonlinear $M(H)$'s far above $T_{M}$ have been
observed in some particular quasi-2D ferromagnets (\textit{e.g.} 
La$_{1.2}$Sr$_{1.8}$Mn$_{2}$O$_{7}$), evidence generally points to mesoscopic
phase-separations, rather than dimensionality, as the cause.\cite{cha}

This situation is also unlikely to change with the order of the transition.
The nucleation energy needed in a first-order transition should suppress the
spin alignment at lower fields and make the $\chi$ more $H$-independent. This
has been demonstrated in La$_{0.67}$(Ba$_{x}$Ca$_{1-x}$)$_{0.33}$MnO$_{3}$,
which exhibits a magnetic transition evolving from first-order to second-order 
with increasing $x$.\cite{mou} Indeed, the observed $1/\chi = H/M$ above $T_{M}$ 
has a weaker $H$-dependence at smaller $x$, \textit{i.e.} those with a
first-order transition. With $k_{B}T_{M}/\mu_{0} \approx 30$~T in 
Ru1222Eu, therefore, it would be hard to interpret its nonlinear 
$M(H)$ up to $t =$ 1--2 as simple spin fluctuations.



\subsection{Superparamagnetic Component and Langevin-Function Fits}

An extensive nonlinearity far above $T_{M}$ (\textit{i.e.} a violation of the 
scale theory), however, may occur if magnetic clusters (or strong 
short-range correlations), which are a common phenomena in both 
manganites and cuprates, form. The above free-energy argument will require 
only a cluster size (or the rigid spin-spin correlation-range) of 
$k_{B}T/HM_{0}$ to create a significant nonlinearity. For example, the nonlinear
$M(H)$ in La$_{1.2}$Sr$_{1.8}$Mn$_{2}$O$_{7}$ (a quasi-2D ferromagnet with a 
$T_{M} \approx 126$~K) up to $t \approx 0.27$ (160~K) was
suggested by Chauvet \textit{et al.} to be related to the 7--9~\AA\ clusters
observed by NPD and ESR.\cite{cha,kim} In the case of Ru1222, the magnetic
size of the possible clusters should be $\geq$~50~$\mu_{B}$ around 
2~$T_{M} \approx 130$~K to make a significant nonlinearity below 5~T. Such 
clusters will contain at least 20~Ru ions, based on the estimated moment 
$\mu_{0} \leq$ 3~$\mu_{B}$/Ru, and should be regarded as superparamagnets.
Experimentally, this suggestion, \textit{i.e.} part of Ru1222 is in a
spin-ordered state far above $T_{M}$, is also in agreement with the 
M\"{o}ssbauer data of Ru1222Gd, where an ordered state with a hyperfine field 
$\approx 30$~T has been demonstrated up to $T_{1} \approx 180$~K.\cite{fel}

To make this argument more quantitative, the isothermal $M$-$H$ of the sample is
analyzed using the Langevin function ctnh($\mu H/k_{B}T)-k_{B}T/\mu H$ of 
superparamagnetic particles, where $\mu$ is the
magnetic moment of the particles. This procedure essentially replaces the
spin-correlation function of $\Phi(r_{i},s_{i},r_{j},s_{j})$ by a
step function of $|r_{i}-r_{j}|$: the spins are assumed to be
tightly aligned within a coherence volume but without any directional
correlation at larger distances, where $r_{i}$ and $s_{i}$ are the position
and the spin-orientation, respectively, of the $i$th spin. The short-range
correlations, therefore, are absorbed into the apparent cluster moment $\mu$
and the weaker long-range part is ignored. The function has been routinely used
in analyzing the particle sizes of granular magnets by fitting the
isothermal $M(H)$ as a superposition of weighted Langevin functions. It has
been demonstrated in such cases that the deduced $\mu$ is a good
approximation of the cluster size if the interparticle interactions are
weak, but serves only as a \textit{lower-limit} if the interactions are
strong.\cite{all} It has been argued, in fact, that the main effect of
interparticle interactions is to raise the effective temperature, $T^{\ast}$, 
leading to an underestimation of $\mu$ by a factor of $T/T^{\ast }$.\cite{all} 
The situation is expected to be more complicated near $T_{M}$
of a second-order transition, where the interactions should be a continuing
function of $|r_{i}-r_{j}|$, and the spin alignment should be
described by the scaling models instead. Experimentally, however, the
$M$(350~K) of La$_{0.67}$Pb$_{0.33}$MnO$_{3}$, a manganite with a second-order
transition at $T_{M} = 336$~K, still fits remarkably well with the
function without additional modifications, \textit{i.e.} either a
distribution of $\mu$ or an additional linear term on $H$ (Fig.~\ref{fig:fig3}a).\cite{xueb} 
The deduced cluster size, \textit{i.e.} $\approx 2a$ at $t = 1.05$, seems to 
also be reasonable, at least as a lower limit,
where $a \approx 4$~\AA\ is the distance between adjacent
Mn-spins. The Langevin function, therefore, seems to be a good
phenomenological tool when $t \geq 1.04$, keeping in mind that
the $\mu$ deduced is only a lower limit. In the case of a first-order
transition, larger deviations are expected due to the suppression of $M$ at
low fields. As an example, the $M$ of a La$_{0.67}$Ca$_{0.33}$MnO$_{3}$
ceramic sample, which has a first-order transition at 265~K,\cite{ter} was
measured at 275~K ($t = 1.04$). The $S$-shape $M$-$H$, which is the result of
domain nucleation,\cite{mou} makes the Langevin function a less effective
description (Fig.~\ref{fig:fig3}b). It is interesting to note, however, that the 
$\mu \approx$ 200~$\mu_{B}$/cluster so-deduced is still comparable or smaller
than the 15~\AA\ spin-spin coherence length, \textit{i.e.} 100~Mn/cluster,
observed by small-angle-neutron-scatering.\cite{ter} This demonstrates,
therefore, that the Langevin function should be proper for a qualitative
estimation of the cluster size in Ru1222Eu, at least as its lower limit. 



The average $M$ between the $H$-increase branch and the $H$-decrease branch in
isothermal $M$-$H$ loops of the Ru1222Eu sample was then calculated and taken as
the equilibrium one. The $M$ of the Ru1222Eu sample at 80~K so-obtained
(triangles in Fig.~\ref{fig:fig3}c), however, deviated significantly from the simple
Langevin function (solid line in Fig.~\ref{fig:fig3}c). In particular, the $M$ is a linear
function of $H$ above 2~T. This should not be a surprise since the paramagnetic
background of Eu is expected to be large in any case, and an additional
linear term should be added to the Langevin function. In fact, this
is obvious even in the raw data of both Ru1212 and Ru1222, where the $M$ above
3--4~T was previously reported being a linear function of $H$ with an almost
$T$-independent slope.\cite{fel,but} Our data, therefore, were fitted as 
$\chi_{0}H + m$ $\cdot$ [ctnh($\mu H/k_{B}T)-k_{B}T/\mu H$] (thick dashed lines 
in Fig.~\ref{fig:fig3}c), with an additional fitting parameter of $\chi_{0}$. The fitting is rather 
good over a broad temperature range (Fig.~\ref{fig:fig4}). It is interesting to note two unusual 
characteristics of the deduced parameters (Fig.~\ref{fig:fig5}). First, the $\mu$ is large and 
decreases with $T$ rather slowly over a broad range of $T \leq$ 150~K (the fitting 
uncertainty makes the situation at higher temperatures less clear), which is very
different from the $\mu$ of critical fluctuations (as suggested by the Ni data
in the same figure).\cite{see} The deduced $\mu$ of Ru1222Ru, in fact, is several
orders of magnitude larger than that of Ni when $T/T_{M} > 1.1$
(open diamonds in Fig.~\ref{fig:fig5}). Such a large $\mu$, \textit{e.g.} 
$\approx$~300~$\mu_{B}$ at $T/T_{M} = 2$, is difficult to understand without a 
cluster formation (or a short-range spin-spin correlation strong enough to prevail
against the thermal disturbance $k_{B}T$), and justified the use of the 
Langevin function. Second, the deduced $m$, which is roughly equal
to the ``extrapolated zero-field moment'' in previous reports, varies
with $T/T_{M}$ only gradually, without any sharp transition-like features
across $T_{M}$ (inset, Fig.~\ref{fig:fig5}). For instance, although the major increase of 
$m$ occurs around 100--120~K, its long tail extends up to $T_{1} \approx$ 
160--180~K. It should also be pointed out that the $m$
deduced does not saturate with cooling down to 10~K and is far smaller than
the aligned moment of 1--2 $\mu_{B}$/Ru (50--100 emu/cm$^{3}$) measured by
NMR. This small and $T$-dependent $m$ (\textit{i.e}. the
superparamagnetic component) suggests either a canted AFM ground state or a
phase-separation, \textit{i.e.} only a fraction of Ru-spins are gradually FM-aligned. 
Butera \textit{et al.}, however, have pointed out that the large
canting angle, $\approx$ $20^{\circ}$ in our case, required would be rather
unusual.\cite{but} The simplest scenario, therefore, would be an eventual
precipitation of FM clusters from the paramagnetic matrix. Both the unusual
nonlinear $M(H)$ far above $T_{M}$ and the relatively small $m$ can be
self-consistently interpreted in this simple model.

\subsection{Slow Dynamics Far Above $T_{M}$}

The formation of clusters (or strong short-range spin-spin correlations) can
be verified by the dynamic response of the sample. A slow dynamics should be
characteristic of a magnetic system with energy barriers comparable to the
thermal energy $k_{B}T$. The energy barrier 
($K\cos^{2}\theta - M_{0}H\cos\theta)V_{c}$,\cite{lot} which can be
reduced to $\approx K\cos^{2}\theta V_{c}$ at low fields $H << K/M_{0}$, is 
much smaller than $k_{B}T$ in the case of paramagnets, but much larger in typical
crystalline ferromagnets, where $K$ and $\theta$ are the anisotropy
coefficient and the angle between $H$ and the easy axis, respectively. In
either case, the magnetic response to the changes of $H$ and $T$ will be either
too fast or too slow to be observed in the experimental time-window. An
observable slow dynamics, \textit{i.e.} a time-dependent response between
10 and $10^{6}$~s, implies energy barriers of 20--40 times $k_{B}T$.
In a system of isolated magnetic clusters, this corresponds to the existence
of clusters with a $V_{c}$ on the order of $10^{-18}$--$10^{-21}$~cm$^{3}$
(10--100~\AA\ in diameter) with 
$10^{4}$~erg/cm$^{3} < K < 10^{7}$~erg/cm$^{3}$ (values typical for 
ferromagnets). The dynamic magnetization of the Ru1222Eu, therefore, was
measured above $T_{M}$ to verify the formation of clusters. First, the
sample was cooled to a temperature between 60 and 100~K at zero field. A
field of $H = 5$~Oe was added after the temperature was stabilized for 30 min. 
The $dc$ $M_{ZFC}$ was then continuously measured as a function
of time $t_{1}$, where $t_{1}$ is the time after the field is activated
(Fig.~\ref{fig:fig6}a). A significant logarithmic increase of the magnetization with time
exists over a broad temperature range. In particular, the rate $d\ln M/d\ln t_{1}$ is far 
larger than our experimental uncertainty of $\approx$~0.003 below 90~K. A logarithmic 
relaxation, in principle, may occur if either the related energy barrier 
$K\cos^{2}\theta V_{c}$ has a particular distribution or a strong interparticle interaction
exists.\cite{hag,lot} In both cases, however, the effective barrier (at least its lower limit) 
is $k_{B}T[(d\ln M/d\ln t_{1})^{-1}+\ln (t_{1}/\tau)]$ $\approx$ 
$k_{B}T(d\ln M/d\ln t_{1})^{-1} = 7.4 \cdot 10^{-20}$~J at 80 K,
where $\tau$ is a characteristic time typically $\approx$~$10^{-10}$~s. The $V_{c}$ 
$\approx$ $7.4 \cdot 10^{-20}/\mu_{B}H_{ai}$ (in units of the FM moment involved), 
therefore, will be 700~$\mu_{B}$ with an estimated anisotropy field $H_{ai}$ of 
11~T, \textit{i.e.} the anisotropy field observed in Ru1212Gd along $c$.\cite{but} It is 
interesting to note that the estimated value is in rough agreement with the size of 
$10^{3}$~$\mu _{B}$ from the Langevin-function fit at the same temperature. Second,
the sample was cooled at an $H = 0.08$~Oe (the field inhomogeneity is 
$\pm 0.01$~Oe over the scanning length of 4~cm) from 200~K to a designated
temperature between 70 and 90~K with different cooling rates between 0.1 and 
$\approx$~50~K/min (by mechanically dropping the sample into a precooled
chamber). The $M_{FC}$ under such conditions was then measured at 30-min intervals
after the temperature was stabilized. The decrease of $M_{FC}$ with the
cooling rate at temperatures far above $T_{M}$, a slow spin-alignment during
the cooling, is again observed (Fig.~\ref{fig:fig6}b).

It should be pointed out that the above temperature range is where two
additional transitions appear, and that part of the relaxation may be associated
with the corresponding species. However, the large $d\ln M/d\ln t_{1}$, as well as its 
smooth $T$-dependence below $T_{M}$, where the
additional species should be negligible, suggests that there should be a 
significant superparamagnetic component for the 65-K species as well. This
is also supported by the slow $ac$ dynamics in Ru1222Eu reported
previously.\cite{hir}



\subsection{Non-Curie-Weiss Susceptibility and the Separated AFM Transition}

It should be noted that the linear term $\chi_{0}$ so-deduced from the
Langevin-function fit, which should be roughly equal to the differential
susceptibility at 5~T in our case, offers another opportunity to verify the
possible phase separations, \textit{i.e.} a spatially separated matrix, from
the clusters embedded. The deduced $\chi _{0}$ is displayed in Fig.~\ref{fig:fig7} as a
function of temperature. To subtract the contributions of Eu, Ce, and CuO$_{2}$, we 
follow the procedure proposed by Butera \textit{et al.} for Ru1212Eu
and adopted by Williams \textit{et al.} for Ru1222Eu, \textit{i.e.} using
the Van Vleck susceptibility of Eu$^{3+}$ and a $T$-independent term of 
$8 \cdot 10^{-6}$ for CuO$_{2}$, and ignoring the contribution of Ce.\cite{but,wil}
It should be pointed out that the Eu/Ce contributions are less certain in
the case of Ru1222. For verification, the same procedure was used for a comparison 
with the measured $\chi$ in the nonmagnetic 
NbSr$_{2}$Eu$_{1.4}$Ce$_{0.6}$Cu$_{2}$O$_{10-\delta}$ (Nb1222Eu), which has 
the same (Eu$_{0.7}$Ce$_{0.3}$)$_{2}$O$_{2}$ block (inset, Fig.~\ref{fig:fig7}). The good 
agreement demonstrates that the procedure is valid and that the contribution of Ce is 
small. In any case, this Eu/Ce/CuO$_{2}$ contribution is relatively small 
($\approx$~$1/2$ of the $\chi_{0}$ observed) and likely to vary with $T$ only 
monotonically (Fig.~\ref{fig:fig7}). The Ru seems to be a major contributor to the $\chi_{0}$ 
(a similar conclusion has been reached previously\cite{but,wil}). This suggestion is
supported by the facts that the raw $\chi_{0}$ of Ru1222Eu exhibits a 
$T$-dependence very different from that of 
NbSr$_{2}$Eu$_{1.4}$Ce$_{0.6}$Cu$_{2}$O$_{10-\delta}$ and the Ru contribution 
to the Curie constant, $\approx$~2.6~$\mu_{B}$/Ru after this
background-subtraction procedure, seems to be reasonable. A large amount of
paramagnetic Ru-spins, therefore, should coexist with the ferromagnetic
clusters in the sample over a broad temperature range. A likely scenario,
therefore, appears to be a phase separation between the FM clusters and a
paramagnetic matrix, although alternate interpretations, \textit{i.e.}
peculiar spin canting or magnetic anisotropy, should also be considered (see
discussions below).



To further verify the phase-separation proposed, $dc$ and differential $ac$ 
susceptibilities up to 400~K were measured and analyzed. Above $T_{1}$~$\approx$ 
180~K, no $H$-dependence of $\chi$ can be noticed; the $M_{FC}$(1~T)$/H$ and 
the $M_{FC}$(5~T)$/H$ merge there (Fig.~\ref{fig:fig2}). Ru1222Eu, or at least its 
dominant part, should be in a simple paramagnetic state above $T_{1}$. The 
C-W law of $\chi = C/(T-T_{CW})$ was then used to fit the data after both 
the Eu-contribution and a $T$-independent term of $8 \cdot 10^{-6}$ were subtracted 
from the raw data. The $1/\chi = H/M_{FC}$(5~T) observed between 
$T_{1} \approx$ 200~K and 400~K is a linear function of $T$ within our 
experimental uncertainty of 0.1--1\% (inset, Fig.~\ref{fig:fig8}). The deduced Ru moment of 
2.61~$\mu_{B}$ is in good agreement with that of 
RuSr$_{2}$EuCeCu$_{2}$O$_{10-\delta}$ previously reported.\cite{wil} The Curie
temperature $T_{CW} \approx$ 80~K is only slightly higher than the $T_{M}$
defined from the low-$H$ $M_{FC}$ (Figs.~\ref{fig:fig1},\ref{fig:fig8}). Below 180~K, 
however, the 
$H/M_{FC}$(1~T) is significantly lower than the $1/\chi$ expected from the C-W
fit, but the differential susceptibility at 5~T is much higher. Such deviation
at so high a $t$ is rather unusual. Theoretically, it has been suggested that the 
C-W law should be applicable down to the critical region if the spin interactions have 
an infinite range.\cite{arr} Even in the case of short-range 3D Ising or Heisenberg 
interactions, the agreement is typically good down to $t \approx$ 
$10^{-1}$.\cite{arr,see} The corrections expected are at a high power of $T_{M}/T$, 
and are typically only 10\% or less when $t > 0.3$. Experimentally, the C-W fit of
Ni is consistent with the data down to $t \approx$ 1.1.\cite{see} Although similar 
deviations have been reported in both spin-glass and manganites, they were attributed to 
cluster formations and/or phase separations.\cite{ter,mor} This can be easily understood. 
On one hand, the cluster formation and phase separation offer a natural interpretation:
the FM clusters will enhance the susceptibility at low fields but suppress the differential 
susceptibility at high fields due to the depletion of the available paramagnetic spins. 
On the other hand, a mean-field treatment leads to a weighted apparent moment of 
$\mu = (3k_{B}/N)^{1/2}/[\frac{d}{dT}(1/\chi)]^{1/2}$.\cite{mor} The gradual increase
(decrease) of $\frac{d}{dT}(1/\chi)$ with cooling, therefore, suggests an increase of 
$\mu$ at low field (high field), \textit{i.e.} cluster formation and phase separation. A 
$\chi$-deviation from the C-W behavior in La$_{2/3}$Ca$_{1/3}$MnO$_{3}$ up to 
2~$T_{M}$, for example, has been taken as evidence for the formation of FM-aligned 
clusters.\cite{ter} The $\chi$ of La$_{1.35}$Sr$_{1.65}$Mn$_{2}$O$_{7}$ is 
even more closely reminiscent of the data in Fig.~\ref{fig:fig8}.\cite{cha} It is $H$-independent 
and follows the C-W law above 380~K with a deduced $T_{CW} \approx$ 280~K. 
Between 280 and 380~K, however, the $H/M$(0.5~T) is far below the C-W fit, but 
the $H/M$(5~T) is far above it. Both NPD and ESR further demonstrate that these 
deviations are the result of phase separations.\cite{cha} The similar $\chi$-behavior of 
Ru1222Eu, therefore, would suggest phase separation and the formation of FM
clusters. It is interesting to further note that the ``missing'' differential susceptibility 
$\Delta \chi_{0,Ru}$ of $C/(T-T_{CW}) - \chi_{0,Ru} \approx$ $5 \cdot 10^{-5}$ at 
120~K is about 30\% of the expected $C/(T-T_{CW}) = 1.65 \cdot 10^{-4}$ 
(Fig.~\ref{fig:fig8}), 
which is only slightly larger than the ratio of $m$(120~K)$/m$(10~K) $\approx$ 20\% 
(inset, Fig.~\ref{fig:fig5}). This is in agreement with the proposed model in which the
homogenous paramagnetic Ru1222 above $T_{1}$ eventually separates into FM
clusters and a matrix on cooling. We further attribute the slight difference between 
these two ratios to local AFM correlations developed in the matrix.
The nonlinear isothermal $M(H)$ and the non-C-W susceptibility up to 2--3~$T_{M}$,
therefore, can be consistently understood. Both are similar to the data
observed in manganites and can be self-consistently interpreted as the
results of the FM clusters resulting from the possible phase separations.



An AFM-like minimum appears around 120~K in the $1/\chi_{0,Ru}$ of the
sample (Fig.~\ref{fig:fig8}). AFM correlations seem to develop quickly below this
temperature. For example, the $\chi_{0,Ru}$ is only 10\% or less of that expected from
the C-W fit below 90~K. It should be pointed out that the 
$m$ at 90~K, \textit{i.e.} the total spins involved in the formation
of the FM-like clusters, remains far smaller than the $m$(10~K) of 
17~emu/cm$^{3}$. The suppression of $\chi_{0,Ru}$, therefore, should
come mainly from the AFM correlations of the matrix. Fisher has proposed, on
rather general grounds, that the magnetic specific heat is propotional to 
$\partial(T\chi)/\partial T$, and that the N\'{e}el temperature corresponds
to the inflection point $\partial^{2}(T\chi)/\partial T^{2} = 0$ in
simple antiferromagnets.\cite{fis} A numerical differential, therefore, was
carried out. The transition temperature $T_{AM} \approx$ 104~K so-deduced
is almost double the $T_{FM} \approx$ 65~K observed (the arrow in
Fig.~\ref{fig:fig8}). It is interesting to note that the $T_{AM}$ differs from the 120-K
dip by only 13\%, characteristic of 3D spin ordering.\cite{de} It should also be
pointed out that the large separation between the $T_{M}$ and the $T_{AM}$ is not
typical of the behavior of homogeneous canted AFM magnets. For example, the 
$M_{FC}$ onset and the peak of the differential susceptibility in CsCoCl$_{3}$, 
a 1D Heisenberg canted antiferromagnet, occur at the same temperature, 
$3.4 \pm 0.2$~K, within the experimental resolution.\cite{her} Ru1222Eu,
therefore, is unlikely to be a simple canted AFM magnet with a single
transition at $T_{M}$. Either several sequential magnetic transitions or a
phase separation should have taken place.

At lower temperatures, an FM-like transition occurs around $T_{M}$ at low
fields (Fig.~\ref{fig:fig1}). It is a puzzle, however, that there are no anomalies in either 
$m$ or $\chi_{0}$ around this temperature, as pointed out
previously.\cite{xue} To explore the nature of the $T_{M}$ transition, the
remnant moment was measured in a $M$-$H$ loop of $\pm 500$~Oe 
(Fig.~\ref{fig:fig9}). It is
interesting to note that there is neither a remnant moment nor a significant
hysteresis above $T_{M}$, although a noticeable $m$ appears there
already---a scenario reminiscent of superparamagnets above their blocking
temperature. This is in agreement with the cluster size deduced in Fig.~\ref{fig:fig5}.
Below $T_{M}$, however, the remnant moment appears, and the $M_{FC}$ at 
100~Oe becomes comparable to the $m$ deduced 
(Figs. \ref{fig:fig3},\ref{fig:fig5},\ref{fig:fig9}). A significant
part of the FM component should possess a magnetic energy 
$HM_{0}V_{c} \geq k_{B}T$ below $T_{M}$, although a superparamagnetic 
component may still exist, as suggested by the slow dynamics. A long-range spin-order,
therefore, is established at $T_{M}$ due either to a rapid growth of the
cluster size or to a phase-coherent transition among adjacent clusters.



The proposed model, in our opinion, can also offer possible explanations for
many conflicting and confusing data previously reported. The conflict between
the NPD and NMR data for the magnetic structure, for example, may be
attributed to the fact that the two probes have different sensitivities to
various magnetic species. The NPD, on one hand, is insensitive to the
minor FM nanoclusters due to both the volume fraction, estimated to be 
$\approx$~10\% for Ru1212Gd, and the submicron size of the clusters. The
sensitivity of NMR (as a $rf$ transverse susceptibility), on the
other hand, is small for AFM (inversely proportional to the exchange field),
large for a single-domain FM (inversely proportional to the anisotropy
field), and even larger for FM domain-walls (proportional to the wall
mobility). The NMR data of Ru1212/Ru1222, therefore, may be dominated by the
minor FM part due to, for example, the small in-plane anisotropy field of 
$\approx$~200~Oe observed.\cite{but} In fact, a similar (though not as dramatic)
situation has been reported in La$_{0.35}$Ca$_{0.65}$MnO$_{3}$.
When the zero-field NMR signal of the compound is dominated by the minor FM
species with a volume fraction of $\approx$~8\%,\cite{kap} its NPD data
show a homogeneous AFM structure.\cite{rad} The different spin orientations
assigned by NPD (\textit{i.e.}, along the $c$-axis) and by NMR 
(\textit{i.e.}, perpendicular to the $c$-axis) can be naturally understood as well.
Similarly, the suppression of the superconducting order parameters inside
the FM nanoclusters and the possible tunneling across them again offer a
natural mechanism for the granular superconductivity observed.\cite{xuea}

Two alternate interpretations have been proposed.\cite{fel,but} It has been 
suggested that either Ru1222 is already in a homogeneous canted AFM state
below $T_{1}$ (a multi-transition model) or that the FM components along the 
$c$- and the $ab$-axes depend on $H$ very differently (a
two-stage spin-alignment model). In such models, however, several features
are expected. a) An anomaly of $\chi_{0,Ru}$ as well as an $m$ jump
are expected around $T_{1}$, where a bulk transition occurs. In fact, the 
$\chi_{0,Ru}$ is determined by the C-W law above $T_{1}$, but by the magnetic
anisotropy and/or the cant angle below $T_{1}$ in such 
models. Similarly, $m$ should be zero above $T_{1}$ but should be determined by the canted 
angle, $\theta$, below $T_{1}$. These two totally different mechanisms make the smooth 
crossovers
around $T_{1}$ purely coincidental. Experimentally, no discontinuity or anomaly
has ever been reported around $T_{1}$, although both $T_{1}$ and $T_{CW}$
can be tuned significantly through doping and oxidation.\cite{xue,fel} b)
The $m$ and the $\chi_{0,Ru}$ should be related between $T_{1}$ and 
$T_{M}$ through $\theta$ and/or the magnetic anisotropy. Their very different
$T$-dependences in Figs. \ref{fig:fig5} and \ref{fig:fig8}, therefore, will require rather peculiar
temperature dependences of the $\theta$ and/or the magnetic anisotropy. c)
The FM-like transition at $T_{M}$ as well as the AFM-like one at $T_{AM}$
will be ``extra'' features in such models.

The situation in Ru1212Eu is slightly different. However, we believe that similar
arguments apply.

\section{Summary}

Nonlinear $M(H)$, non-C-W behavior, and slow spin dynamics have been
observed in a Ru1222Eu sample between $T_{M} = 65$~K and $T_{1} = 180$~K,
suggesting the formation of superparamagnetic clusters $10^{2}$--$10^{3}$ $\mu _{B}$ 
in size. Additionally, the differential susceptibility $\chi_{0,Ru}$ shows an AFM transition at 
$T_{AM}$ $\approx$ 104~K. We
therefore propose a phase-separation model: the sample is a mesoscopic
mixture of FM clusters and a paramagnetic matrix between $T_{1}$ and $T_{M}$,
followed by an AFM transition of the matrix at $T_{AM}$ and a long-range
phase-coherent transition in the imbedded clusters at $T_{M}$. This model
offers a possible interpretation for many conflicting data, but it needs to be
verified by further structure investigations.

\begin{acknowledgments}
We thank I. Felner for helpful discussions and for the
unpublished Nb1222Eu susceptibility data. The work in Houston is supported in part by 
NSF Grant No. DMR-9804325, the T. L. L. Temple Foundation, the John J. and Rebecca 
Moores Endowment, and the State of Texas through the Texas Center for Superconductivity 
at the University of Houston; and at Lawrence Berkeley Laboratory by the Director, Office of 
Science, Office of Basic Energy Sciences, Division of Materials Sciences and Engineering of the 
U.S. Department of Energy under Contract No. DE-AC03-76SF00098.
\end{acknowledgments}


\begin{figure}
\caption{\label{fig:fig1}$M/H$ of the 
RuSr$_{2}$(Eu$_{0.7}$Ce$_{0.3}$)$_{2}$Cu$_{2}$O$_{10+\delta}$ 
sample at 10~Oe. $\bullet$: ZFC; $\bigcirc$: FC.
The arrow indicates the inflection point $T_{M}$ of the $M_{FC}(T)$.}
\caption{\label{fig:fig2}$\chi = M_{FC}/H$ of several compounds. Solid 
symbols: Ru1222Eu with $H = 0.0005$, 0.01, 0.1, 1, 2, and 5~T from top to 
bottom.}
\caption{\label{fig:fig3}$M$ vs $H$ for: a) 
La$_{0.67}$Pb$_{0.33}$MnO$_{3}$ at 350~K ($t = 1.05$); b) 
La$_{0.67}$Ca$_{0.33}$MnO$_{3}$ at 275~K ($t = 1.04$); and c) Ru1222Eu at 
80~K ($t = 1.27$).
Symbols: data; ---: fit to the Langevin function; - - -: fit to the Langevin
function with an additional linear term.}
\caption{\label{fig:fig4}$M$ vs $H$ for Ru1222Eu  at 70, 80, 90, 100, 110, 
120, and 130~K from top to bottom. Symbols: data; ---: fit to the Langevin function
with a linear term (see text).}
\caption{\label{fig:fig5}The estimated cluster size in the Ru1222Eu sample ($\bullet$) 
and in Ni ($\Diamond$).\cite{see} Inset: the saturation moment $m$ of the proposed 
magnetic species in Ru1222Eu.}
\caption{\label{fig:fig6}a) The relaxation of the $M_{ZFC}$ at 5~Oe. b) The 
$M_{FC}$ at 0.08~Oe with different cooling rates. Measurements were done at 
the temperatures marked on the right side.}
\caption{\label{fig:fig7}The extracted linear term $\chi_{0}$ ($\bullet$) of Ru1222Eu 
and the estimated contribution from Eu$^{3+}$ and CuO$_{2}$ (---). Inset: ---: 
the estimated contribution from Eu$^{3+}$ and CuO$_{2}$ to 
NbSr$_{2}$Eu$_{1.4}$Ce$_{0.6}$Cu$_{2}$O$_{10+\delta}$; $\bullet$: data 
(courtesy of I. Felner).}
\caption{\label{fig:fig8}The $T$-dependence of $1/\chi$ due to Ru in Ru1222Eu. 
$\bigtriangleup$: $H/M_{FC}$(1~T); $\bigtriangledown$: $H/M_{FC}$(5~T); 
$\square$: differential susceptibility at 5~T; ---: the Curie-Weiss fit. Inset: $1/\chi$ 
between 200 and 400~K. $\bullet:$ data; ---: the C-W fit.}
\caption{\label{fig:fig9}The remnant magnetization of Ru1222Eu after a $\pm 500$~Oe 
field cycle at various temperatures.}
\end{figure}


\end{document}